\documentclass[12pt,preprint]{aastex}

\usepackage{graphicx}

\begin{document}

\title{Efficient Simulations of Interstellar 
Gas-Grain Chemistry \\ 
Using Moment Equations}
\author{Baruch Barzel and Ofer Biham} 
\affil{ 
Racah Institute of Physics,  
The Hebrew University,  
Jerusalem 91904,  
Israel} 

\begin{abstract} 

Networks of reactions on dust grain surfaces
play a crucial role in the chemistry of interstellar clouds,
leading to the formation of molecular hydrogen in diffuse 
clouds as well as various organic molecules in dense
molecular clouds.
Due to the sub-micron size of the grains
and the low flux, the population of reactive
species per grain may be very small and strongly fluctuating.
Under these conditions rate equations fail and
the simulation of surface-reaction networks requires stochastic methods 
such as the master equation.
However, the master equation becomes infeasible 
for complex networks because
the number of equations proliferates exponentially.
Here we introduce a method based on moment equations
for the simulation of reaction networks on small grains.
The number of equations is reduced
to just one equation per reactive specie and one equation
per reaction. 
Nevertheless, the method provides accurate results, which
are in excellent agreement with the master equation.
The method is demonstrated for the methanol network  
which has been recently shown to be of crucial importance.

\end{abstract} 


\section{Introduction}

Chemical networks in interstellar clouds consist of  
gas-phase and grain-surface reactions
\citep{Hartquist1995,Tielens2005}.
Reactions that take place on dust grains include the formation
of molecular hydrogen 
\citep{Gould1963,Hollenbach1971b}
as well as reaction networks producing ice
mantles and various organic molecules.
Unlike gas phase reactions in cold clouds that mainly produce 
unsaturated molecules, surface processes  
are dominated by hydrogen-addition reactions
that result in saturated, hydrogen-rich molecules, such as
H$_2$CO, CH$_3$OH, NH$_3$ and CH$_4$.
In particular, recent experiments show that methanol cannot
be efficiently produced by gas phase reactions
\citep{Geppert2006}.
On the other hand, there are indications that it can be 
efficiently produced on ice-coated grains
\citep{Watanabe2005}. 
Therefore, the ability  
to perform simulations of the production of
methanol and other complex molecules on grains is of great importance
\citep{Garrod2006}. 
Unlike gas-phase reactions,   
simulated using rate equation models
\citep{Pickles1977,Hasegawa1992}, 
grain-surface reactions require stochastic methods such as 
the master equation
\citep{Biham2001,Green2001}, 
or Monte Carlo (MC) simulations 
\citep{Charnley2001}. 
This is due to the fact that under interstellar conditions,
of extremely low gas density and sub-micron grain sizes, 
surface reaction rates are dominated by fluctuations
which cannot be accounted for by rate equations 
\citep{Tielens1982,Charnley1997,Caselli1998,Shalabiea1998}. 
A significant advantage of 
the master equation over MC simulations is that
it consists of differential equations, which
can be easily coupled to the 
rate equations of gas-phase chemistry. 
Furthermore, unlike MC simulations that require the accumulation
of statistical information over long times, the master equation
provides the probability distribution from which the reaction rates
can be obtained directly.
However, the number of equations increases exponentially
with the number of reactive species,
making the simulation of complex networks infeasible 
\citep{Stantcheva2002,Stantcheva2003}. 
The recently proposed multi-plane method
dramatically reduces the number of equations, 
by breaking the network into a set of fully connected 
sub-networks 
\citep{Lipshtat2004},
enabling the simulation of more complex networks. 
However, the construction of the multi-plane 
equations for large networks turns out to be 
difficult.
 
In this Letter we introduce a method based on moment equations 
which exhibits crucial advantages over the multi-plane method.
The number of equations is further reduced to the smallest
possible set of 
stochastic equations,
including one equation for the population size of each reactive specie
(represented by a first moment) 
and one equation for each reaction rate (represented by a second moment). 
Thus, for typical sparse networks the complexity of the stochastic 
simulation becomes comparable to that of the rate 
equations.
Unlike the master equation (and the multi-plane method) 
there is no need to adjust the cutoffs -
the same set of equations applies under all physical conditions.
Unlike the multi-plane equations, the moment equations are
linear and for steady state conditions can be easily solved using
algebraic methods.
Moreover, for any given network the moment equations
can be easily constructed using a diagrammatic approach, which can 
be automated
\cite{Barzel2007}.

\section{The Method}

\begin{figure} 
\includegraphics[width=5in]{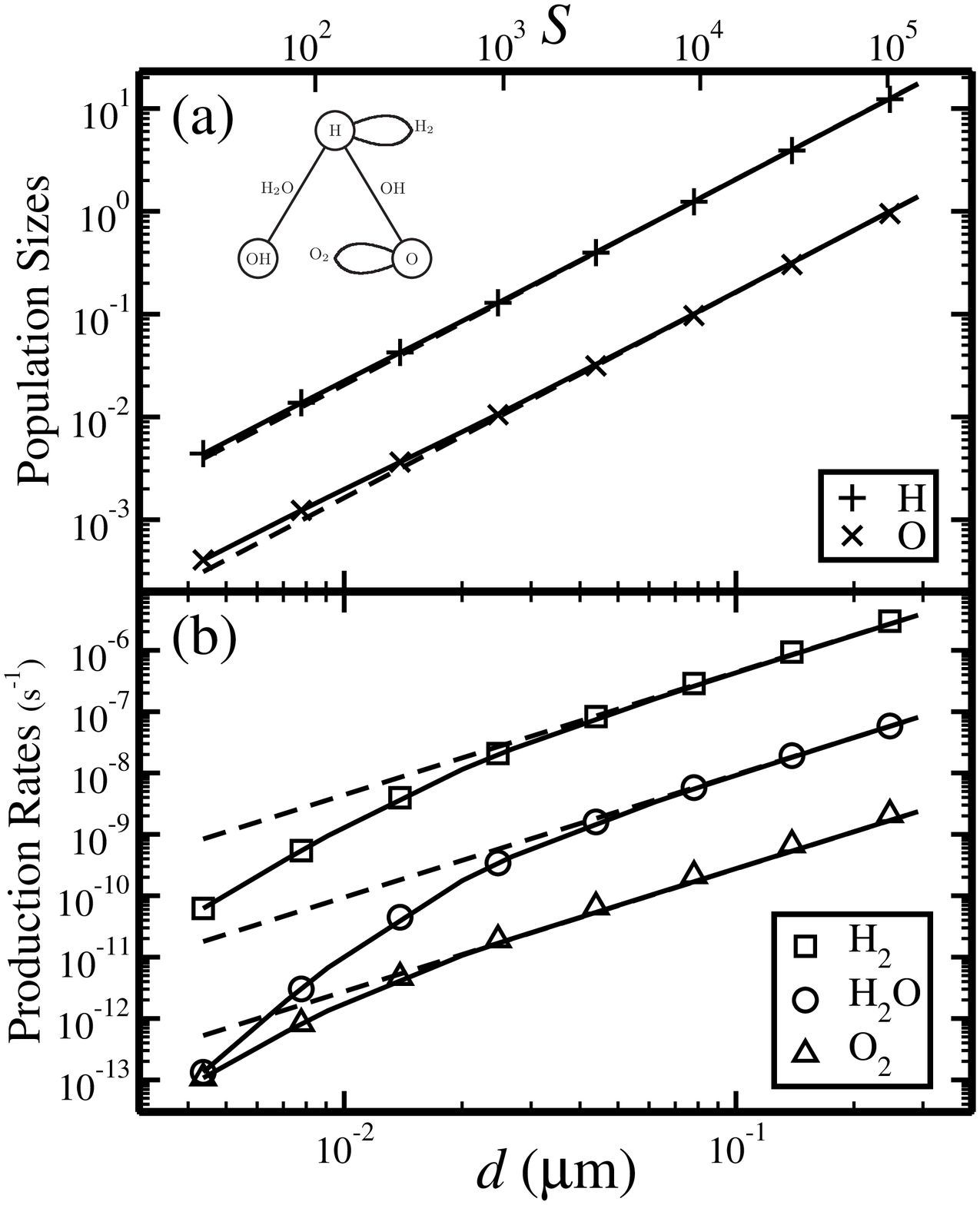}
\caption{ 
(a) The population sizes of H ($+$) and O ($\times$) atoms on a grain
vs. grain diameter, $d$ (and the number of adsorption sites, $S$), 
obtained from the moment equations for the
network shown above, where nodes represent reactive species, edges
represent reactions and the products are indicated near the
edges; 
(b)
The production rates
of H$_2$ (squares),
O$_2$ (triangles) 
and 
H$_2$O (circles) 
on a grain 
vs. grain 
diameter,
obtained from the moment equations. 
The results are
in excellent agreement with the
master equation (solid line).
In the limit of large grains they also coincide with 
the rate equations (dashed line).
\label{fig:1} 
} 
\end{figure}

To demonstrate the method we consider a simple network, 
shown in Fig. 
\ref{fig:1},
that 
involves three reactive species: H and O atoms and OH molecules 
\citep{Caselli1998,Shalabiea1998,Stantcheva2002}. 
For simplicity we denote the reactive species by 
$X_1=$ H, 
$X_2=$ O, 
$X_3=$ OH, 
and the non-reactive product species
by 
$X_4=$ H$_2$, 
$X_5=$ O$_2$, 
$X_6=$ H$_2$O. 
The reactions that take place in this network include 
H + O
$\rightarrow$
OH 
($X_1 + X_2 \rightarrow X_3$), 
H + H
$\rightarrow$
H$_2$ 
($X_1 + X_1 \rightarrow X_4$), 
O + O
$\rightarrow$
O$_2$ 
($X_2 + X_2 \rightarrow X_5$), 
and 
H + OH
$\rightarrow$
H$_2$O 
($X_1 + X_3 \rightarrow X_6$). 
 
Consider a spherical grain of diameter $d$, 
exposed to fluxes 
of H and O atoms and OH molecules. 
The cross-section of the grain is
$\sigma=\pi d^2/4$
and its surface area is $\pi d^2$.
The density of adsorption sites on the surface
is denoted by
$s$ (sites cm$^{-2}$).
Thus, the number of 
adsorption sites on the grain is
$S=\pi d^2 s$.
The desorption rates of atomic and molecular species from the 
grain are given by  
$W_i =  \nu \cdot \exp [- E_{1}(i) / k_{B} T]$,   
where $\nu$ is the attempt rate  
(standardly taken to be $10^{12}$ s$^{-1}$),  
$E_{1}(i)$  
is the activation energy for desorption  
of specie 
$X_i$ 
and $T$ (K) 
is the grain temperature. 
The hopping rate of adsorbed atoms between
adjacent sites on the surface is
$a_i =  \nu \cdot \exp [- E_{0}(i) / k_{B} T]$,  
where 
$E_{0}(i)$ is the activation energy for hopping
of $X_i$ atoms (or molecules). 
Here we assume that diffusion occurs only by thermal hopping, 
in agreement with experimental results 
\citep{Katz1999,Perets2005}. 
For small grains it is convenient to 
define the scanning rate,
$A_i = a_i / S$,
which is approximately the inverse of the time it takes 
an $X_i$ atom to 
scan the surface of the entire grain.

The master equation provides the time
derivatives of the probabilities 
$P(N_1,N_2,N_3)$ 
that on a random grain there will be $N_i$ adsorbed atoms/molecules
of the reactive specie $X_i$. 
It takes the form 
 
\begin{eqnarray} 
&&\dot P(N_1,N_2,N_3) =
\sum_{i=1}^3 
F_i \left[ P(..,N_i-1,..) - P(N_1,N_2,N_3) \right] 
\nonumber \\
&&+\sum_{i=1}^3 
W_i \left[ (N_i+1) P(..,N_i+1,..) - N_i P(N_1,N_2,N_3) \right] 
\nonumber \\
&&+ \sum_{i=1}^2 
A_i [ (N_i+2)(N_i+1) P(..,N_i+2,..)
- N_i(N_i-1) P(N_1,N_2,N_3) ] 
\nonumber\\
&&+ (A_1+A_2) [ (N_1+1)(N_2+1) P(N_1+1,N_2+1,N_3-1)
- N_1 N_2 P(N_1,N_2,N_3) ]
\nonumber \\
&&+ (A_1+A_3) [ (N_1+1)(N_3+1) P(N_1+1,N_2,N_3+1)
- N_1 N_3 P(N_1,N_2,N_3) ].
\label{eq:Master} 
\end{eqnarray} 

\noindent 
The terms in the first sum describe the incoming flux, 
where $F_i$ (atoms s$^{-1}$)  
is the flux {\em per grain} 
of the specie $X_i$. 
The second sum describes the effect of desorption.  
The third sum describes the effect of diffusion mediated reactions
between two atoms of the same specie 
and the last two terms account for reactions between different species. 
The rate of each reaction is proportional 
to the number of pairs of atoms/molecules 
of the two species involved, and to the sum of their scanning rates. 
The moments of
$P(N_1,N_2,N_3)$
are given by
$\langle N_1^a N_2^b N_3^c \rangle = 
\sum_{N_1,N_2,N_3}
N_1^aN_2^bN_3^c P(N_1,N_2,N_3)$, 
where $a,b,c$ are integers.  
In particular, 
$\langle N_{i}\rangle$ 
is the average population size   
of the specie $X_i$ on a grain.
The production rate
per grain,
$R(X_k)$
(molecules s$^{-1}$), 
of $X_k$ molecules 
produced by the reaction 
$X_i + X_j \rightarrow X_k$ 
is given by
$R(X_k) = (A_i+A_j) \langle N_i \  N_j \rangle$, 
or by 
$R(X_k)=A_i \langle N_i(N_i-1) \rangle$ 
in case that $i=j$. 
 
In numerical simulations the master equation
must be truncated in order to
keep the number of equations finite. 
This can be done by setting 
upper cutoffs 
$N_i^{\max}$, $i=1,\dots,J$
on the population sizes, 
where $J$ is the number of reactive species.
However, the number of coupled equations, 
$N_E = \prod_{i=1}^J (N_i^{\max}+1)$, 
grows exponentially
with the number of reactive species. 
This severely limits the applicability of the master equation 
to interstellar chemistry 
\citep{Stantcheva2003}. 
To reduce the number of equations one tries to use the lowest 
possible cutoffs under the given conditions.
In any case, to enable all reaction processes to take place, 
the cutoffs must satisfy 
$N_i^{\rm max} \ge 2$
for
species that form homonuclear diatomic molecules 
(H$_2$,O$_2$, etc.)
and 
$N_i^{\rm max} \ge 1$ 
for other species.
 
The average population sizes of the reactive species
and the reaction rates 
are completely determined by all the first moments 
and selected second moments
of the
distribution 
$P(N_1,N_2,N_3)$. 
Therefore, a closed set of equations for the time derivatives of 
these first and second moments could provide complete information 
on the population sizes and reaction rates. 
For the simple network considered here one needs equations 
for the time derivatives of the first moments 
$\langle N_1 \rangle$, 
$\langle N_2 \rangle$
and 
$\langle N_3 \rangle$ 
and of the second moments 
$\langle N_1^2 \rangle$, 
$\langle N_2^2 \rangle$, 
$\langle N_1 N_2 \rangle$ 
and 
$\langle N_1 N_3 \rangle$ 
[nodes and edges, respectively, in the graph shown in Fig. \ref{fig:1}].
Such equations are obtained 
by taking the time derivative of each moment
and using Eq.
(\ref{eq:Master})
to express
the time derivatives of the probabilities
\citep{Lipshtat2003}.
Here we show two of the resulting moment equations: 
 
\begin{eqnarray} 
\frac{d \langle N_{1} \rangle }{dt} &=& 
F_{1}+(2A_{1}-W_{1}) \langle N_{1} \rangle
-2A_{1} \langle N_{1}^2 \rangle 
-(A_{1}+A_{2}) \langle N_{1}N_{2} \rangle
-(A_{1}+A_{3}) \langle N_{1}N_{3} \rangle
\nonumber 
\\ 
\frac{d \langle N_{1}N_{3} \rangle }{dt} &=& 
F_{1} \langle N_{3} \rangle +F_{3} \langle N_{1} \rangle
-(W_{1}+W_{3}-3A_{1}-A_{3}) \langle N_{1}N_{3} \rangle
-(3A_{1}+A_{3}) \langle N_{1}^2N_{3} \rangle
\nonumber\\ 
&-&(A_{1}+A_{3}) \langle N_{1}N_{3}^2 \rangle
+(A_{1}+A_{2}) (\langle N_{1}^2N_{2} \rangle
- \langle N_{1}N_{2} \rangle
- \langle N_{1}N_{2}N_{3} \rangle).
\label{eq:moment1} 
\end{eqnarray} 
 
\noindent 
In these equations, 
the time derivative of each moment 
is expressed as a linear combination of several other moments.
However, the right hand sides of these equations include 
third order moments for which we have no equations. 
In order to close the set of moment equations 
we must express the third order moments in terms of
first and second order moments 
\citep{Lipshtat2003}.
This can be done by imposing the following constraint
on the master equation: 
at any given time, at most two atoms or molecules can be 
adsorbed simultaneously on the surface. Furthermore, 
these two atoms or molecules must be from species that react 
with each other. 
The resulting cutoffs allow only eight non-vanishing probabilities, 
namely, 
$P(0,0,0)$, 
$P(0,0,1)$, 
$P(0,1,0)$, 
$P(1,0,0)$, 
$P(2,0,0)$, 
$P(0,2,0)$, 
$P(1,1,0)$ 
and 
$P(1,0,1)$. 
The third moments in Eq. 
(\ref{eq:moment1})
can now be expressed 
in terms of these non-vanishing probabilities, giving rise to
the following rules:
(a) $\langle N_1 N_2 N_3 \rangle = 0$; 
(b) $\langle N_i^2 N_j \rangle = \langle N_i N_j \rangle$; 
(c) $\langle N_i^3 \rangle = 
3 \langle N_i^2 \rangle - 2 \langle N_i \rangle$. 
Using these rules,
which are general and apply to any network of binary reactions,
one can modify Eqs. 
(\ref{eq:moment1}), 
and obtain a closed set of the form 
 
\begin{eqnarray} 
\frac{d \langle N_{i} \rangle }{dt} &=& 
F_{i}-W_{i} \langle N_{i} \rangle
-2A_{i} (\langle N_{i}^2 \rangle - \langle N_{i} \rangle ) 
-(A_{1}+A_{2}) \langle N_{1}N_{2} \rangle
- \delta_{i,1} (A_{i}+A_{3}) \langle N_{i}N_{3} \rangle
\nonumber 
\\ 
\frac{d \langle N_{3} \rangle }{dt} &=& 
F_{3} - W_{3} \langle N_{3} \rangle
-(A_{1}+A_{3}) \langle N_{1}N_{3} \rangle 
+(A_{1}+A_{2}) \langle N_{1}N_{2} \rangle 
\nonumber 
\\ 
\frac{d \langle N_{i}^2 \rangle }{dt} &=& 
F_{i}+(2F_{i}+W_{i}+4A_{i}) \langle N_{i} \rangle
-(2W_{i} +4A_{i}) \langle N_{i}^2 \rangle  
\nonumber\\ 
&-& 
(A_{1}+A_{2}) \langle N_{1}N_{2} \rangle
- \delta_{i,1} (A_{i}+A_{3}) \langle N_{i}N_{3} \rangle
\nonumber 
\\ 
\frac{d \langle N_{1}N_{j} \rangle }{dt} &=& 
F_{1} \langle N_{j} \rangle +F_{j} \langle N_{1} \rangle
-(W_{1}+W_{j}+A_{1}+A_{j}) \langle N_{1}N_{j} \rangle,
\label{eq:moment2} 
\end{eqnarray} 
 
\noindent 
where $i=1,2$, $j=2,3$ and $\delta_{i,j}=1$ if $i=j$ and 0 otherwise.
This set includes one equation that accounts for 
the population size of each reactive species 
and one equation that accounts for the rate 
of each reaction. 
Although these equations were derived using strict 
cutoffs, that are expected to apply only in the limit of 
very small grains and low flux,
they provide accurate results for a very broad range of 
conditions.
The point is that once the set of moment equations is
derived, the probabilities do not appear anymore,
so the constraint is not explicitly enforced.
In fact, the equations maintain their accuracy even when 
the populations sizes of the reactive species are well beyond 
the constraints imposed above.

In Fig.
\ref{fig:1}(a) 
we present the population sizes of 
H ($+$) 
and 
O ($\times$)
atoms on a grain,
vs. grain diameter,
obtained from the moment equations.
In Fig.
\ref{fig:1}(b) 
we present the production rates of 
H$_2$ (squares),
O$_2$ (triangles) 
and 
H$_2$O (circles) 
vs. grain diameter,
obtained from the moment equations. 
The results are in excellent agreement with the master 
equation (solid lines). In the limit of large grains they also 
coincide with the rate equations (dashed lines).
Note that the moment equations apply even when there
are as many as 10 hydrogen atoms on a grain. 
The parameters used in the simulations are 
$s=5 \times 10^{13}$ (sites cm$^{-2}$), 
$F_1 = 5.0 \times 10^{-10} S$ (atoms s$^{-1}$), 
$F_2 = 0.1 F_1$
and 
$F_3=0$. 
The activation energies for diffusion and desorption were taken as 
$E_0(1) = 44$, $E_1(1)=52$, $E_0(2) = 47$, 
$E_1(2)=54$, $E_0(3)=47$ and $E_1(3)=54$ meV.
The grain temperature was $T=15$K. 
The parameters used for hydrogen are the 
experimental results for low density amorphous ice
\citep{Perets2005}. 
For the other species there are no concrete experimental results, 
and the values reflect the tendency of heavier species to bind 
more strongly. 
The fluxes and grain temperatures are suitable for dense molecular 
clouds.
 
\section{The Methanol Network}

\begin{figure} 
\includegraphics[width=5in]{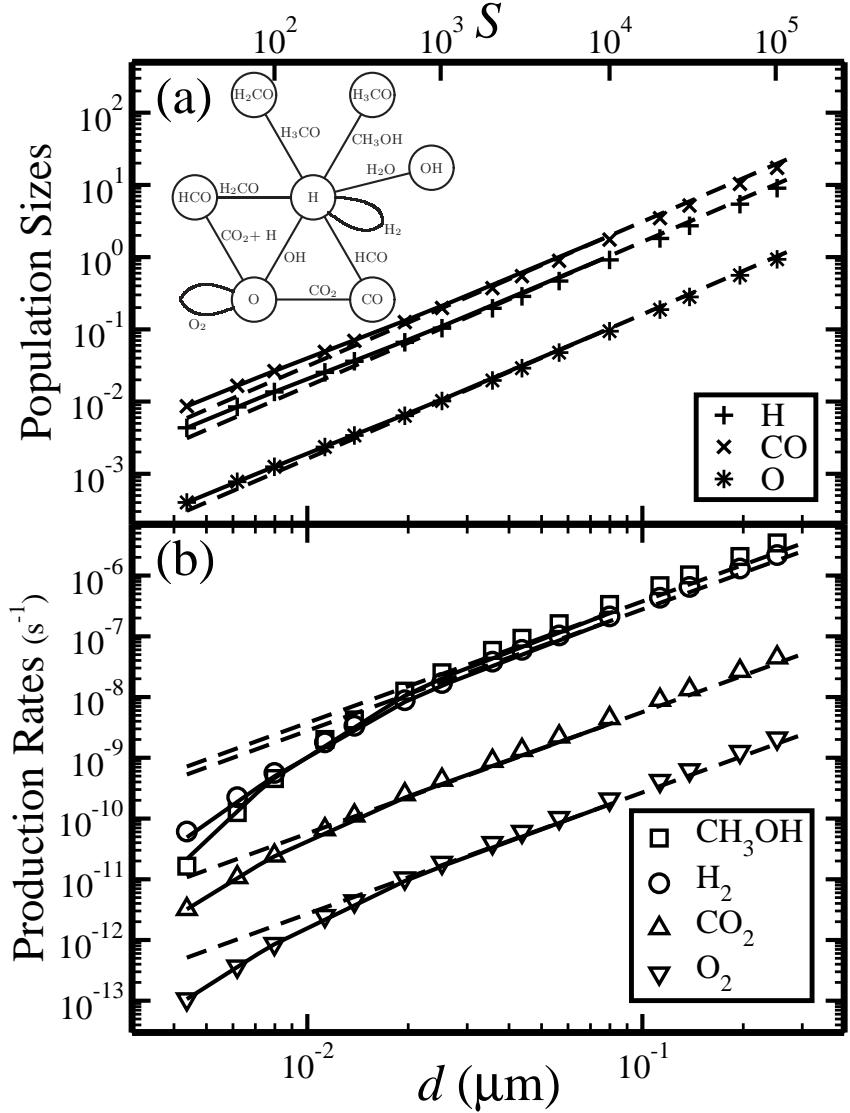}
\caption{ 
(a) The population sizes of 
H ($+$), O ($\ast$) and CO ($\times$)
for the methanol
network shown above, 
vs. grain diameter, $d$ (and the number of adsorption sites, $S$),
obtained from the moment equations. 
(b)
The production rates
of several final products of the methanol network
per grain
vs. grain 
diameter.
obtained from the moment equations. 
The results are
in excellent agreement with the
master equation (solid line).
In the limit of large grains they also coincide with 
the rate equations (dashed line).
\label{fig:2} 
} 
\end{figure}

Consider the case in which a flux of CO molecules 
is added to the network.
This gives rise to the
network shown in
Fig. 
\ref{fig:2},
which includes the 
following sequence of hydrogen addition 
reactions 
\citep{Stantcheva2002}: 
H + CO $\rightarrow$ HCO, 
H + HCO $\rightarrow$ H$_2$CO, 
H + H$_2$CO $\rightarrow$ H$_3$CO
and 
H + H$_3$CO $\rightarrow$ CH$_3$OH. 
Two other reactions that involve oxygen atoms also take place: 
O + CO $\rightarrow$ CO$_2$ 
and
O + HCO $\rightarrow$ CO$_2$ + H. 
This network was studied before using the multiplane method, 
which required about a thousand equations compared to about a 
million equations in the master equation with similar cutoffs
\citep{Lipshtat2004}.
The moment equations include one equation for each node and 
one for each edge, namely, the network shown in
Fig. 
\ref{fig:2} requires only 17 equations. 
We have performed extensive simulations of this network
using the moment equations and found that they are in 
excellent agreement with the master equation.
In Fig. \ref{fig:2}(a) we present the moment-equation results 
for the population sizes of H, O and CO on a grain vs. grain
diameter for the methanol network.
In Fig. \ref{fig:2}(b) we present the moment-equation results 
for the production rates per grain of some of the final products of the
network.
The results are in excellent agreement with the master equation
(solid lines), and coincide with the rate equations 
(dashed lines) for large grains.
The activation energies for diffusion and desorption 
of CO ($X_7$), HCO ($X_8$), H$_2$CO ($X_9$) and H$_3$CO ($X_{10}$) 
were taken as 
$E_0(7) = 50 $, 
$E_1(7) = 55 $, 
$E_0(8) = 52 $, 
$E_1(8) = 58 $, 
$E_0(9) = 53 $, 
$E_1(9) = 59 $, 
$E_0(10) = 55 $
and 
$E_1(10) = 62 $ meV. 
The flux of CO molecules was taken as 
$F_7 = 0.2 F_1$. 
Note that experiments on CO desorption from ice surfaces indicate
that $E_1(7)$ should be slightly higher than the value used here
\citep{Collings2004}.
However, it turns out that using a higher value would compromise the
feasibility of the master equation simulations,
which are required here
in order to establish the validity of the moment equation method.

When simulating highly complex networks one encounters 
the problem of obtaining the equations themselves. 
The ability to automate the construction of the equations 
becomes important. 
A crucial advantage of the moment equations 
is that they can be easily obtained 
using a diagrammatic approach.
A detailed presentation of the diagrammatic method will be
given in 
Barzel \& Biham (2007).
 
\section{Summary and Discussion}

In summary, we have introduced a method, based on moment equations, 
for the simulation of chemical networks
taking place on dust-grain surfaces 
in interstellar clouds.
The method provides highly efficient simulations of complex reaction 
networks under the extreme conditions of low gas density 
and sub-micron grain sizes, in which the reaction rates 
are dominated by fluctuations and stochastic simulations
are required.
The number of equations is reduced to  
one equation for each 
reactive specie and one equation for each reaction,
which is the lowest possible number for such networks. 
This method enables us to efficiently
simulate networks of any required complexity
without compromising the accuracy. 
It thus becomes possible to incorporate
the complete network of surface reactions 
into gas-grain models of interstellar chemistry.
To fully utilize the potential of this method,
further laboratory experiments are needed, that
will provide the
activation energy barriers for diffusion, desorption
and reaction processes not only for hydrogen
but for all the molecules involved in these networks.

We thank A. Lipshtat for helpful discussions.
This work was supported by the Israel Science Foundation and the
Adler Foundation for Space Research.

\clearpage 
\newpage 
 
\end{document}